\documentclass[aps,pra,twocolumn,superscriptaddress]{revtex4}

\usepackage{graphicx}           
\usepackage{amsmath,amssymb}
\usepackage{textcomp}

\begin{document}

\bibliographystyle{naturemag}

\title{Room-Temperature Quantum Memory for Polarization States}

\author{Connor Kupchak, Thomas Mittiga, Bertus Jordaan, Mehdi Namazi, Christian N{\"o}lleke and Eden Figueroa}
\affiliation{Department of Physics and Astronomy, Stony Brook University, New York 11794-3800, USA}

\begin{abstract}
An optical quantum memory is a stationary device that is capable of storing and recreating photonic qubits with a higher fidelity than any classical device. Thus far, these two requirements have been fulfilled in systems based on cold atoms and cryogenically cooled crystals. Here, we report a room-temperature quantum memory capable of storing arbitrary polarization qubits with a signal-to-background ratio higher than 1 and an average fidelity clearly surpassing the classical limit for weak laser pulses containing 1.6 photons on average. Our results prove that a common vapor cell can reach the low background noise levels necessary for quantum memory operation, and propels atomic-vapor systems to a level of quantum functionality akin to other quantum information processing architectures.
\end{abstract}

\maketitle

A readily available, technologically simple, and inexpensive platform for optical quantum memories is the cornerstone of many future quantum technologies \cite{Kimble2008,Lvovsky2009,Bussieres2013,Northup2014}. The practical implementation of such devices is fundamental to realizing deterministic logic gates for optical quantum computing \cite{Fan2012,Monroe2014}, and creating quantum repeater stations that overcome the current distance-limits of quantum key distribution \cite{DLCZ}. A robust and truly scalable architecture may benefit from room-temperature, easy-to-operate quantum light-matter interfaces. Despite much progress \cite{Maurer2012,Saeedi2013,Eisaman2005, Reim2011, Hosseini2011_2, Sprague2014}, the storage of qubits in a room-temperature system has not yet been demonstrated \cite{Novikova2011}.

Room-temperature systems have shown much promise towards advanced optical technologies with progressions such as the miniaturization of vapor cells \cite{Baluktsian2010} and their integration into photonic structures for applications like light slow down \cite{Ghosh2006}, four-wave mixing \cite{Londero2009}, cross-phase modulation \cite{Venkataraman2013} and storage \cite{Sprague2014}.  Furthermore, a warm vapor alleviates the need for laser trapping and cooling in vacuum or cooling to cryogenic temperatures.

The storage of light in atomic vapor can operate with high efficiency (87\%) \cite{Hosseini2011_1}, large spectral bandwidth (1.5 GHz) \cite{Reim2010} and storage times on the order of milliseconds \cite{Novikova2011}.  Vapor systems have proven their ability to preserve non-classical properties in the storage and retrieval of quantum light states \cite{Novikova2011}.  In regard to qubits, polarization states were shown to be stored with high fidelity in experiments involving bright light pulses \cite{Cho2010,England2012}.  However, complete quantum memory operation [i.e. storage of polarization qubits] using warm atomic vapors has yet to be achieved \cite{Novikova2011} due to large control-field-related background photons constraining the \emph{signal-to-background-ratio} (SBR) during retrieval.



\begin{figure*}[]
\centerline{\includegraphics[width=1.4\columnwidth]{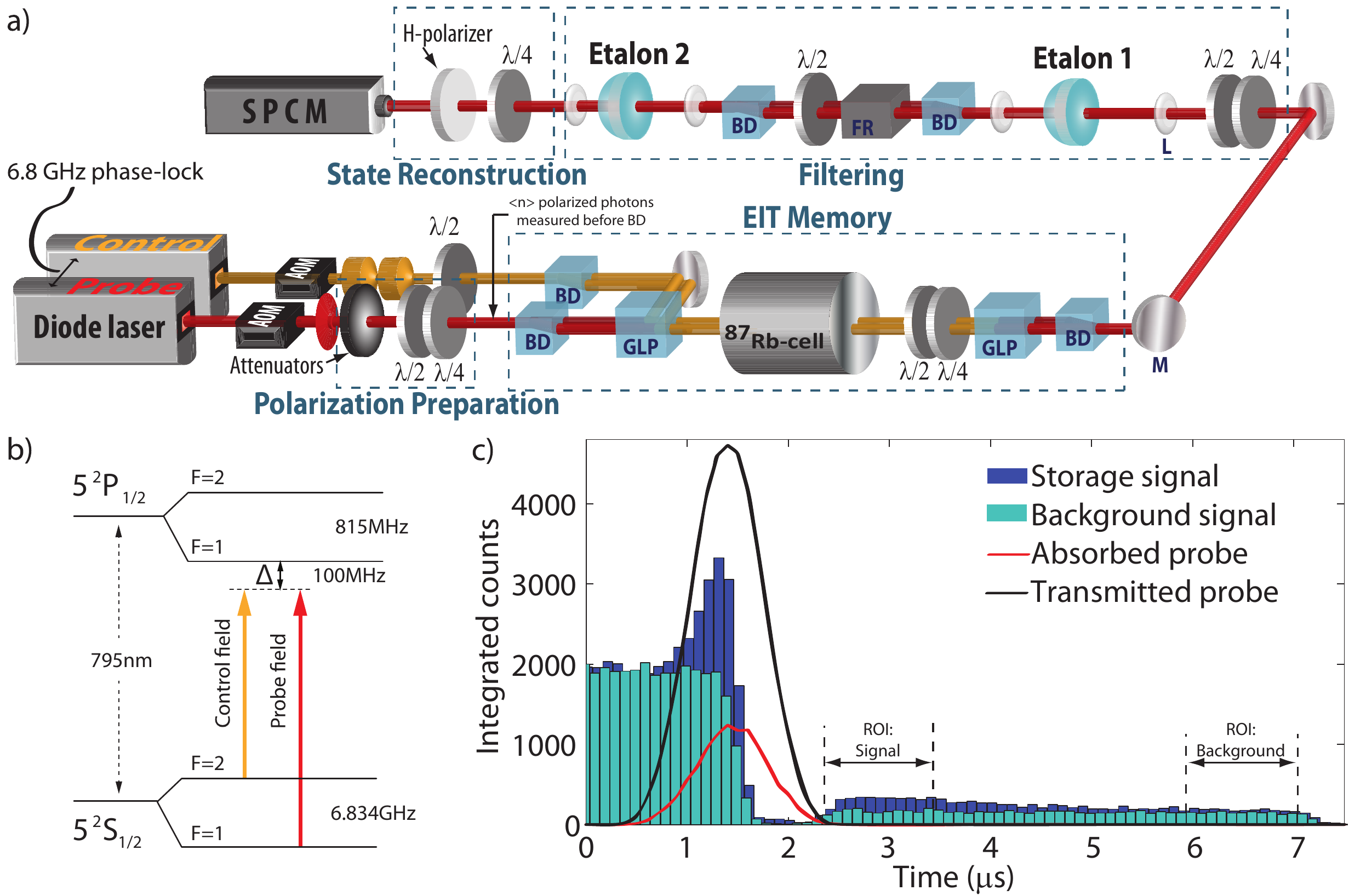}}
\caption{\textbf{Experimental setup and photon-arrival histograms.}  (a) Experimental setup for polarization qubit storage in rubidium vapor at the single-photon level, including the stages of control-filtering. AOM: Acusto-optical modulators; BD: Beam displacers; GLP: Glan-Laser-Polarizer; FR: Faraday rotator; SPCM: Single-Photon-Counting-Module; L: Lens; M: Mirror. Probe: red beam paths; control: yellow beam paths. (b) Atomic level scheme and EIT configuration. (c) Histograms of photon-arrival times, including the input pulse after transmission through the filtering stages (black line), input pulse after absorption in the cell (red line), storage experiment (blue bars) and background (light green bars). The region of interest (ROI) for the data analysis is also displayed.}
\end{figure*}

Here we demonstrate the first room-temperature implementation of an optical quantum memory for qubits, by mapping arbitrary polarization states of light into and out of a warm rubidium vapor. The memory performance is tested with weak coherent pulses containing on average 1.6 photons. The average fidelity is measured to be 71.5 $\pm$ 1.6\%, with qubit coherence times on the order of 20 \textmu s. We also show a detailed analysis of the background noise and its influence on the quantum memory fidelity.

To store a polarization qubit of the form $|\psi_{in}\rangle=$cos$\theta |H\rangle +e^{i\phi}$sin$\theta |V\rangle$ (where $|H\rangle$ and $|V\rangle$ refer to horizontal and vertical polarization states and $\theta$ and $\phi$ correspond to the polar and azimuthal angles on the Poincar\'{e} sphere, respectively), we map the photonic polarization mode onto two spatially separated atomic ensembles concurrently under conditions of electromagnetically-induced transparency (EIT), in a single $^{87}$Rb vapor cell at $62^{\circ}\,$C, containing Ne buffer gas (Figure 1a).

We employed two external-cavity diode lasers phase-locked at 6.8 GHz to resonantly couple a Lambda configuration composed of two hyperfine ground states sharing a common excited state. The probe field frequency is stabilized to the $5S_{1/2} F = 1$ $\rightarrow$ $5P_{1/2} F' = 1$ transition at a wavelength of 795 nm (red detuning $\Delta$=100 MHz) while the control field interacts with the $5S_{1/2} F = 2$ $\rightarrow$ $5P_{1/2} F' = 1$ transition.

The pulse shapes for both the probe and control fields are independently controlled with acousto-optical modulators.  Two polarization beam displacers are used to create a dual-rail set-up allowing simultaneous light-storage in both rails. A set of polarization elements supply 42 dB of control field attenuation while maintaining 80\% probe transmission. Furthermore, two monolithic, temperature-controlled etalon resonators provide a further 102 dB of control field extinction. Both etalons have a thickness of 7.5 mm, radius of curvature of 40.7 mm, free spectral range of 13.3 GHz, finesse of 310 and transmission linewidth of 43 MHz. Together they achieve 16\% probe transmission. In between the etalons we have implemented a polarization insensitive Faraday isolator in order to suppress any back reflections off the etalon surfaces (transmission $\sim$ 50 \%). Overall, our setup achieves 144 dB control field suppression while yielding a total 4.5\% probe field transmission, hence exhibiting an effective, control/probe suppression ratio of 130 dB.

Storage experiments are performed with 1 \textmu s long probe pulses containing 1.6 photons for six different input polarizations ($|H\rangle$, $|V\rangle$, $|D\rangle=\frac{1}{\sqrt2}(|H\rangle+|V\rangle)$, $|A\rangle=\frac{1}{\sqrt2}(|H\rangle-|V\rangle)$, $|R\rangle=\frac{1}{\sqrt2}(|H\rangle +i |V\rangle)$, $|L\rangle=\frac{1}{\sqrt2}(|H\rangle -i |V\rangle)$, forming three mutually unbiased bases of the qubit Hilbert space. The resulting histograms of photon arrival times at the detector contain information regarding both the storage process and events associated to the control-field induced background (\emph{storage histogram}, blue in Fig. 1c).


\begin{figure}[htb]
\centerline{\includegraphics[width=1.0\columnwidth]{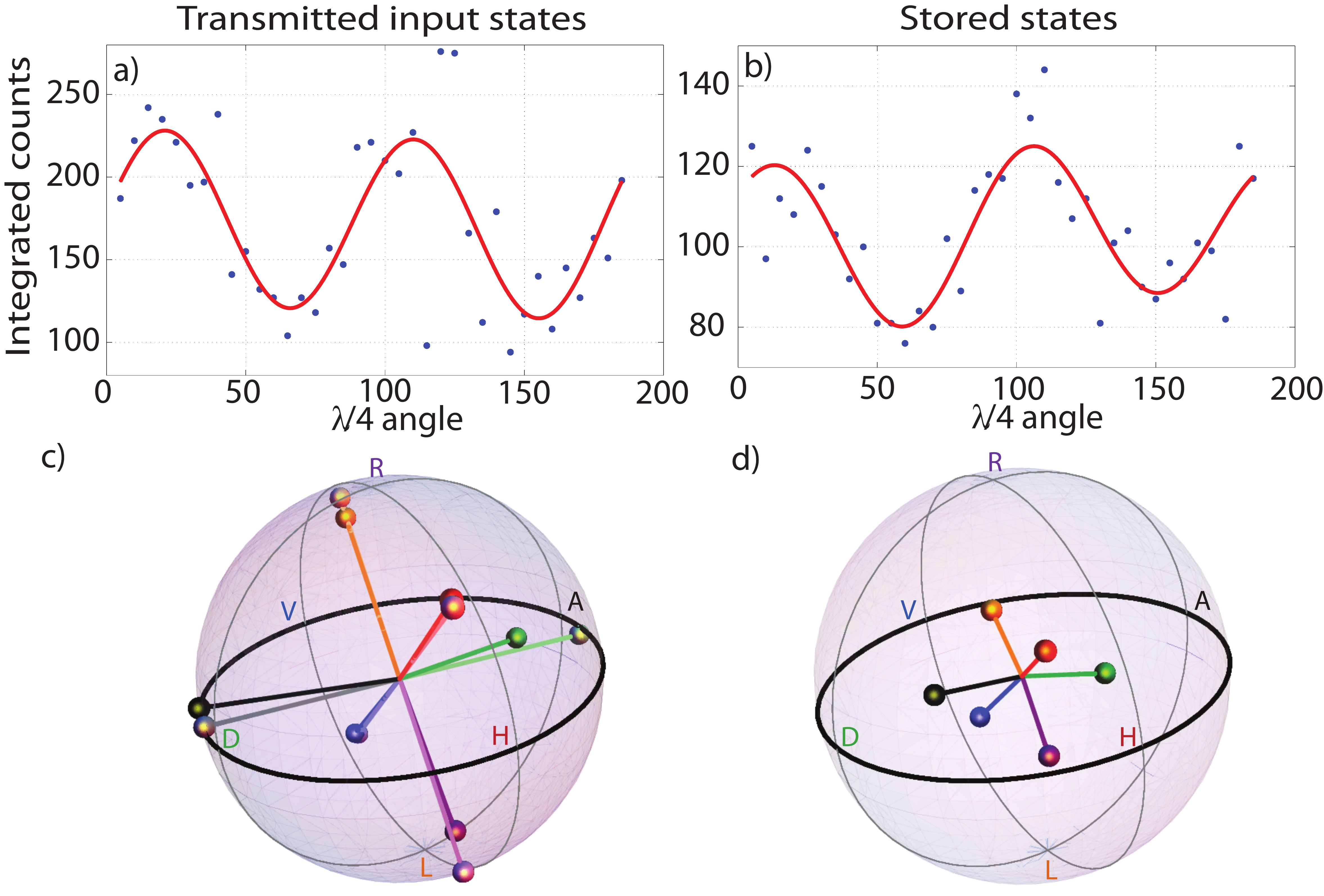}}
\caption{\textbf{Polarization analysis.} Storage of polarization qubits at $\langle n \rangle=1.6$. (a) Stokes reconstruction of $|D\rangle$ transmitted input. (b) Stokes reconstruction of $|D\rangle$  stored and retrieved output. The red line is the fitting used to estimate the Stokes vector. (c) Poincar\'{e} sphere of the transmitted input polarizations (bold colors) and Poincar\'{e} sphere of the rotated input polarizations (light colors). (d) Poincar\'{e} sphere of the stored and retrieved output polarizations.}
\end{figure}

In order to determine the storage efficiency ($\eta$) we integrate the number of counts over the region of interest (ROI) corresponding to the retrieved pulse (from 2.4 to 3.4 \textmu s in Fig. 1c) and subtract the number of counts from a signal-free region of the same histogram (from 6 to 7 \textmu s in Fig. 1c). The efficiency is then calculated by comparing this difference in counts to the total counts in the transmitted probe through the filtering system without atomic interaction (black line in Fig. 1c). The signal to background ratio is obtained in a similar fashion using the counts integrated over the same ROI in the storage histogram (signal+background) and the number of counts over a signal-free region in the same histogram (background). Our SBR is then calculated as [(signal+background)-(background)]/(background) for each polarization input.

The polarization states retrieved from the EIT memory are evaluated using a polarimeter consisting of a quarter-wave plate and polarizer situated after the final filtering stage (see Fig. 1b). Rotating the quarter-wave plate causes oscillations in the intensity measured after the polarizer (within the previously defined ROI), from which we obtain the Stokes vectors ($\textbf{S}=[S_{0},S_{1},S_{2},S_{3}]$, normalized by $S_{0}$) through a fitting (see Fig. 2a-b) \cite{Berry1977}.

The complete evaluation of the polarization fidelity is performed in four steps: First, we measure the Stokes parameters of our input probe polarization entering the first beam displacer. Second, we perform the same procedure for pulses that have propagated through the entire setup (cell included) and the filtering stages in the absence of EIT conditions (see Fig. 1c, red line).  Third, we estimate and apply the unitary rotation to the original input states due to all optical elements by using a least squares fit method which fits them to the transmitted states without changing their lengths (see Fig. 2c).

The fidelity between the rotated inputs ($\textbf{S}_{in}$) and the transmitted states was greater than 99\% on average (green dots in Figure 3a). This step can alternatively be achieved in the system using linear optical elements. Lastly, we perform a polarization analysis of the retrieved pulses ($\textbf{S}_{out}$) which are then compared directly to the rotated input states to obtain a fair estimation of fidelity with respect to the original input states. The fidelity is evaluated as  $F=\frac{1}{2}(1+\textbf{S}_{out}\cdot \textbf{S}_{in}+\sqrt{(1-\textbf{S}_{out}\cdot \textbf{S}_{out})(1-\textbf{S}_{in}\cdot \textbf{S}_{in})})$. We note this procedure is equivalent to utilizing the corresponding density matrices \cite{Altepeter2005}.

In Figure 2d, the Poincar\'{e} sphere associated with the retrieved states clearly shows orthogonal but shortened vectors (as compared to the input) due to the influence of decoherence processes and the uncorrelated background counts. Table 1 summarizes the storage efficiency, SBR, and fidelity reconstruction for all the polarization inputs for $\langle n\rangle=1.6$. We see an average fidelity of 71.5 $\pm$ 1.6\%, clearly surpassing the classical limit of 66\%.


\begin{table}[ht]
\caption{Storage of polarization states in ROI}
\centering
\begin{tabular}{| c | c | c | c | c | c | c | c |}
\hline
\textbf{Input} & \textbf{H} & \textbf{V} & \textbf{D} & \textbf{A} & \textbf{R} & \textbf{L} & \textbf{Average}\\
\hline
SBR & 1.68 & 1.1 & 1.27 & 1.15 & 1.53 & 1.38 & 1.35 $\pm$ .09 \\
\hline
Fidelity (\%)& 71.3 & 79 & 69.2 & 71.4 & 70.2 & 67.6 & 71.5 $\pm$ 1.6 \\
\hline
Efficiency ($\eta$)(\%) & 7.9 & 5.3 & 4.6 & 3.8 & 5.6 & 5.9 & 5.5 $\pm$ .6 \\
\hline
\end{tabular}
\end{table}

To quantify the influence of the background on the fidelity of the qubit memory, we have performed a series of polarization measurements (using the ROI as before), where we modify the SBR by increasing the input photon number (see Figure 3a). We can see that an average fidelity of 90\% can be achieved at a SBR of $\sim$ 8.0. The scaling of SBR can be understood with a theoretical model considering a dual-rail optical quantum memory based on two atomic ensembles, with each ensemble assumed to be a Poissonian source of uncorrelated signal and background photons.

We assume that each of the ensembles stores one of the two polarization components with efficiency $\eta$ before recombination and read out. The probability of producing $n$ signal photons and $m$ background photons (for both ensembles) is $P'_s(n)=\frac{(\eta p)^n}{n!}e^{-\eta p}$ and $P'_{bg}(m)=\frac{q^m}{m!}e^{-q}$ respectively.  Here $p$ is the average number of input photons, and $q$ is the average number of background photons. Note that two ensembles emitting Poissonian noise with mean photon number $q/2$ into the same spatial mode behave as one noise source with mean photon number $q$.\\
In the instance of $n$ signal and $m$ background photons being produced, the probability of detecting a signal photon is simply $\frac{n}{n+m}$, and of detecting a background photon is $\frac{m}{n+m}$ for non photon-number resolving detectors. Then, in general, the probabilities of detecting up to order $N$ signal $P_s(\eta,p,q,N)$ and background $P_{bg}(\eta,p,q,N)$ photons are
$$P_s(\eta,p,q,N)=\sum_{n=0}^N\sum_{m=0}^N P'_s(n)P'_{bg}(m)\frac{n}{n+m}$$
$$P_{bg}(\eta,p,q,N)=\sum_{n=0}^N\sum_{m=0}^N P'_s(n)P'_{bg}(m)\frac{m}{n+m},$$
and the fidelity is
$$F=\frac{P_s(\eta,p,q,N)+\frac{1}{2} P_{bg}(\eta,p,q,N)}{P_s(\eta,p,q,N)+P_{bg}(\eta,p,q,N)}.$$
The theoretical estimation for the fidelity scaling (solid red line in Fig. 3a) has been calculated using independently measured parameters $\eta$=0.055 and  $q=0.005$ (see Figs. 4a-b).

Additionally, we have also measured the coherence time of the qubit storage. Figure. 3b shows the decay time of the quantum memory for the case $\langle n \rangle$=6, showing a $1/e$ time of 19.3 \textmu s.


\begin{figure}[]
\centerline{\includegraphics[width=1.0\columnwidth]{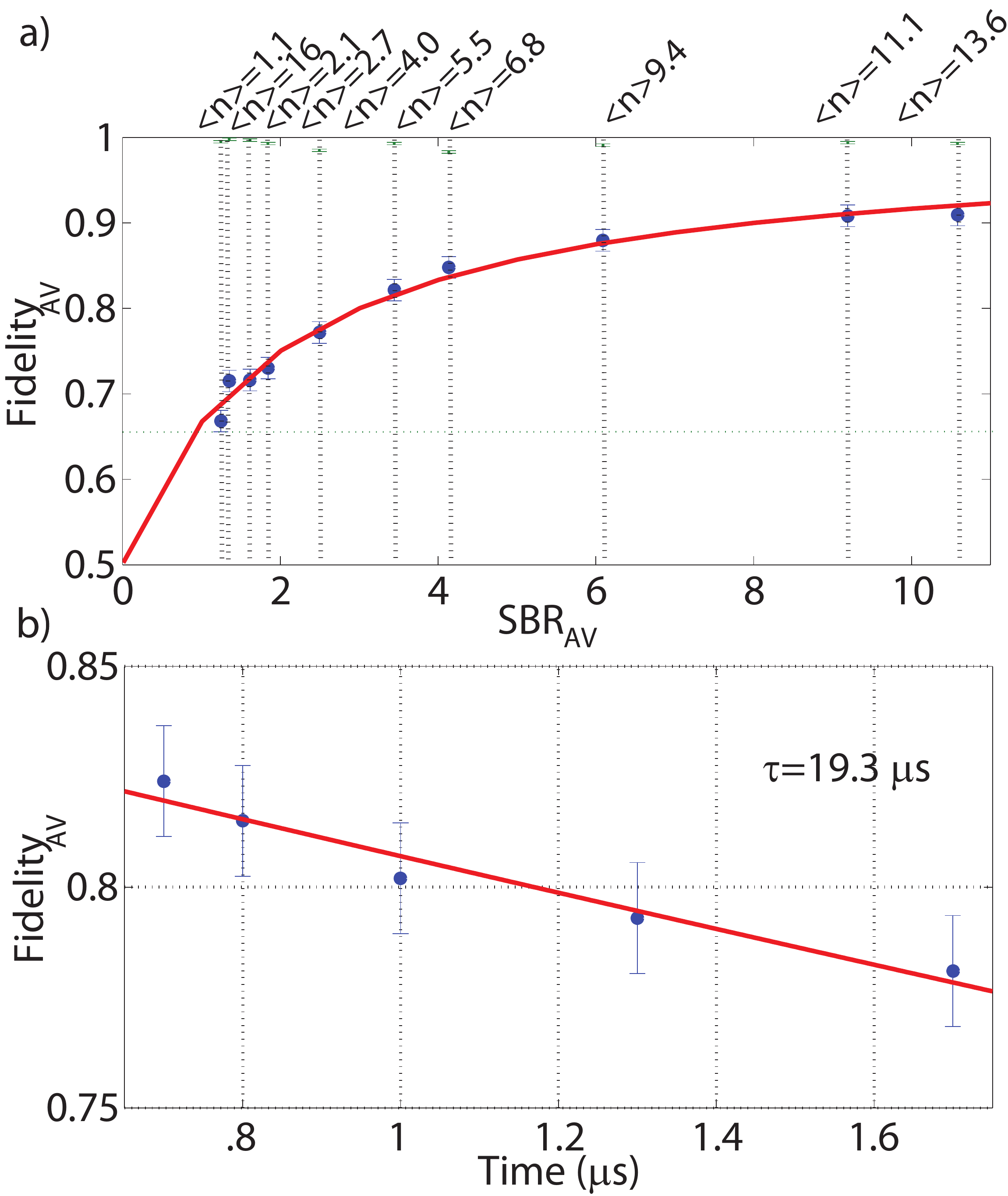}}
\caption{\textbf{Analysis of the quantum memory.} (a) Scaling of the average fidelity of the qubit memory for varying signal-to-background ratio (transmitted states: green dots, retrieved states: blue dots). Shown in black are the average input photon numbers and the corresponding average signal-to background ratio. The red line shows the results of a theoretical model considering a dual-rail optical quantum memory, assuming each ensemble to be a Poissonian source of uncorrelated signal and background photons. (b) Coherence time measurement for $\langle n \rangle$=6 (blue dots) and life-time fitting (red line). The error bars in the measurements are statistical.}
\end{figure}

\begin{figure}[htb]
\centerline{\includegraphics[width=1.0\columnwidth]{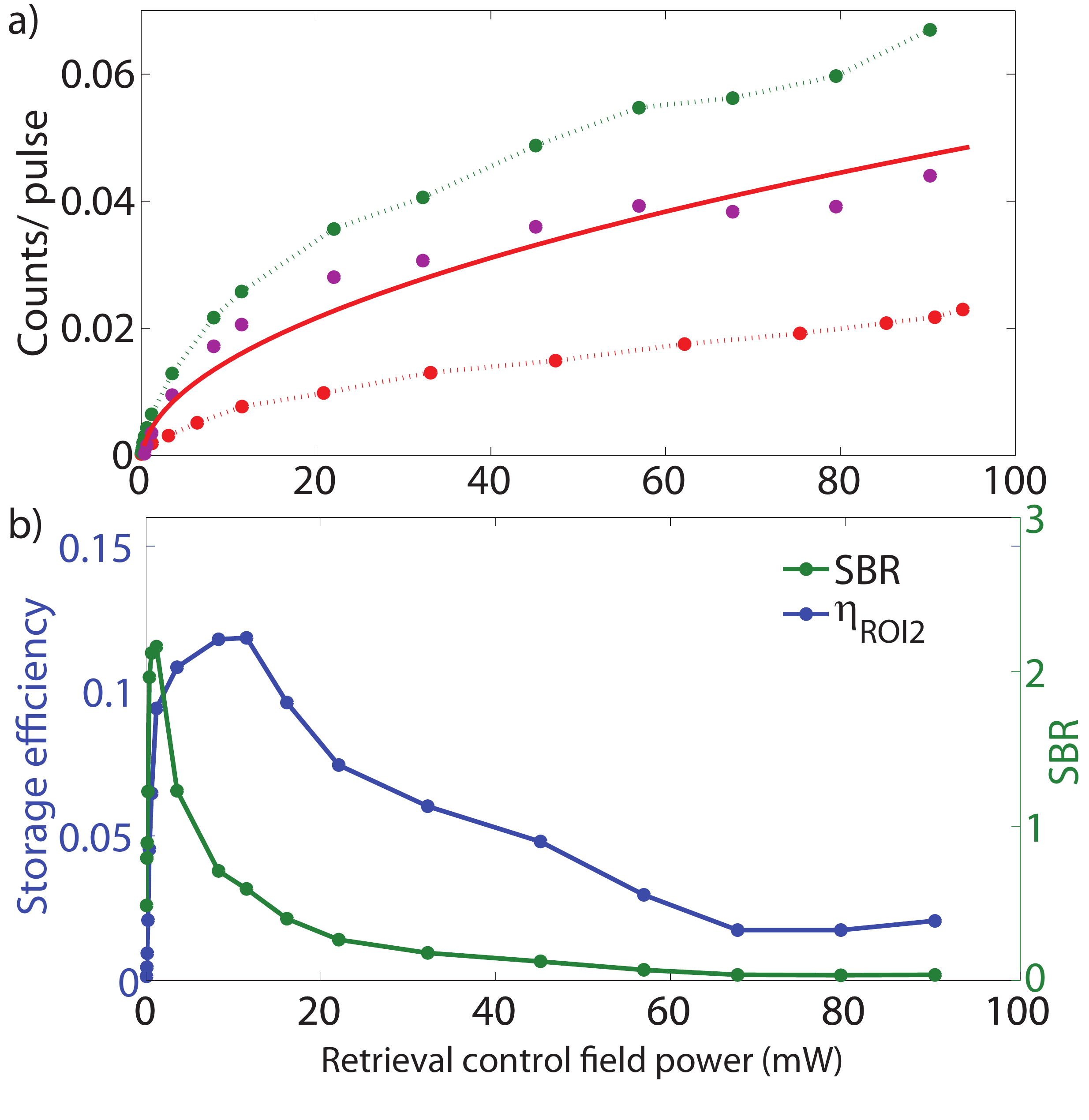}}
\caption{\textbf{Background noise characterization.} (a) Counts in ROI per retrieved pulse for background (green dots) and technical background (control field only, no cell, red dots) with increasing control field power. The purple dots show the background counts with the technical counts subtracted and the red line is a fitting of a function $\propto \sqrt{POWER_{\Omega_{c}}}$. (b) Storage efficiency in the ROI (blue dots) and signal to background (green dots) as a function of control field power.}
\end{figure}

Furthermore, we experimentally characterize the background noise. To do so we integrate the number of counts in the ROI of histograms corresponding to measurements of only the background (cell present, control field only, green dots in Fig. 4a) and only the technical background (control field only, no cell, red dots in Fig. 4a) and divide by the number of experimental runs. This provides the number of background counts per retrieved pulse. We repeat this procedure for several values of the control field power.

We can see that the total background is composed of photons from both leakage of the control field (technical background) and those generated by spontaneous emission and four-wave mixing (FWM) processes \cite{Phillips2011,Lauk2013}. Background photons due to FWM originate from unprepared atoms in the storage region that are pumped to the F=1 ground state by the control field during retrieval. The number of resulting FWM photons should then scale as $\propto \Omega_{c}$ (or the square root of the control field power), where $ \Omega_{c}$ is the Rabi frequency of the control field. The purple dots in Fig. 4a show the resultant of the technical counts subtracted from the background and the red line is a fitting of a function $\propto \sqrt{POWER_{\Omega_{c}}}$, suggesting that our background is dominated by the the FWM mechanism.

Lastly, we analyze the behaviour of the storage efficiency in the ROI ($\eta$, blue dots in Fig. 4b) and SBR (using the same ROI, green points in Fig. 4b) as a function of the control field power. We can see that the efficiency has a substantially different scaling than the SBR and that their maxima do not match. We notice that while our setup is capable of maximum storage efficiencies of $\eta_{max} \sim$ 16\% (over a larger ROI), the ideal signal to background value for quantum memory functionality corresponds to suboptimal storage efficiencies.

In summary, we have presented the first, to our knowledge, room-temperature optical quantum memory system capable of storing arbitrary polarization states. We have demonstrated an average fidelity of 71.5 $\pm$ 1.6\% and storage lifetimes of $\sim$ 20 micro-seconds. We have also investigated the influence of the background in the fidelity of the qubit memory and provided a model explaining the scaling of fidelity with signal-to-background ratio. These measurements demonstrate that a six-fold decrease in background is still necessary for our current implementation to operate simultaneously at higher fidelities and with maximum efficiencies. This could be achieved by using an additional re-pumper scheme \cite{Jang2006} or by modifying the one-photon detuning of the laser system. Longer coherence times can be achieved by adding paraffin coating to our current cells \cite{Balabas2010}. We believe that the present system has the potential to be implemented on a grand scale and thus paves the way for the creation of novel quantum repeaters and networks based on truly scalable architectures.
\newline

We thank A. I. Lvovsky and A. J. MacRae for sharing their etalon design. This work was supported by the College of Arts and Sciences and the Office of the Vice President for Research of Stony Brook University. The authors kindly thank G. Rempe, S. Ritter, A. Neuzner, J. Shupp and A. Reiserer for useful discussions.

\end{document}